\documentclass[apjl]{emulateapj}
\usepackage{amssymb,amsmath,amsbsy}
\usepackage{booktabs}
\usepackage{multirow}
\usepackage{url}

\newcommand{\given}{\,|\,}

\newcommand{\Msun}{\ifmmode {{\rm M}_{\odot}}\else M$_{\odot}$\fi}
\newcommand{\bs}[1]{\boldsymbol{#1}}
\newcommand{\degree}{^{\circ}}
\newcommand{\eqn}{Equation~}

\newcommand{\period}{T}
\newcommand{\mf}{m_f}
\newcommand{\wdupper}{1.44}

\begin{document}

\title{The mass distribution of companions to low-mass white dwarfs}
\author{Jeff J.~Andrews\altaffilmark{\colum}, Adrian M.~Price-Whelan\altaffilmark{\colum}, Marcel A.~Ag\"ueros\altaffilmark{\colum}}

% Affiliations
\newcommand{\colum}{1}
\altaffiltext{\colum}{Department of Astronomy, 
		              Columbia University, 
		              550 W 120th St., 
		              New York, NY 10027, USA}

\begin{abstract}
Measuring the masses of companions to single-line spectroscopic binary stars is (in general) not possible because of the unknown orbital plane inclination.  Even when the mass of the visible star can be measured, only a lower limit can be placed on the mass of the unseen companion. However, since these inclination angles should be isotropically distributed, for a large enough, unbiased sample, the  companion mass distribution can be deconvolved from the distribution of observables. In this work, we construct a hierarchical probabilistic model to infer properties of unseen companion stars given observations of the orbital period and projected radial velocity of the primary star. We apply this model to three mock samples of low-mass white dwarfs (LMWDs, $M\lesssim0.45~\Msun$) and a sample of post-common-envelope binaries. We use a mixture of two Gaussians to model the WD and neutron star (NS) companion mass distributions. Our model successfully recovers the initial parameters of these test data sets. We then apply our model to 55 WDs in the extremely low-mass (ELM) WD Survey. Our maximum a posteriori model for the WD companion population has a mean mass $\mu_{\rm WD} = 0.74~\Msun$, with a standard deviation $\sigma_{\rm WD} = 0.24~\Msun$. Our model constrains the NS companion fraction $f_{\rm NS}$ to be $<$16\% at 68\% confidence. We make samples from the posterior distribution publicly available so that future observational efforts may compute the NS probability for newly discovered LMWDs.
\end{abstract}

\keywords{binaries: general --- binaries: spectroscopic --- methods: statistical --- white dwarfs} 

\section{Introduction}
Except in cases of extreme metallicity \citep{kilic07}, the Galaxy is not old enough to produce low-mass white dwarfs (LMWDs) through single-star evolution. Instead, LMWDs are expected to form through interactions with another star \citep{han98,nelemans00,nelemans01,vdSluys06,woods12}. Indeed, with few exceptions, follow-up observations consistently find companions to LMWDs \citep{marsh95,maxted00,nelemans05,rebassa11}. Recently, the ELM WD Survey has identified 61 extremely LMWDs ($M\lesssim0.3~ \Msun$) in the Sloan Digital Sky Survey \citep[SDSS;][]{york00} and elsewhere \citep{ELMI,ELMII, ELMIII, ELMIV, ELMV}. We refer to the 55 WDs found by these authors that have a measured radial velocity (RV) and orbital period ($\period$) as the ELM sample.

These RV and $\period$ measurements indicate that the LMWDs companions are most likely WDs. However, since the inclination angle $i$ is unknown, LMWDs could have neutron star (NS) companions. Indeed, LMWDs are known companions to millisecond pulsars, although these WDs are generally too faint for spectroscopy \citep{vKerkwijk96,callanan98,bassa06,antoniadis12}. Finding even one NS companion to a spectroscopically characterized LMWD would be very valuable, since this system could constrain the NS mass. To date, unfortunately, radio and X-ray searches for NS companions to LMWDs have been unsuccessful \citep{vLeeuwen07,agueros09b,agueros09a,kilic13}. 

\begin{figure}[h!]
\begin{center}
\includegraphics[width=0.95\columnwidth]{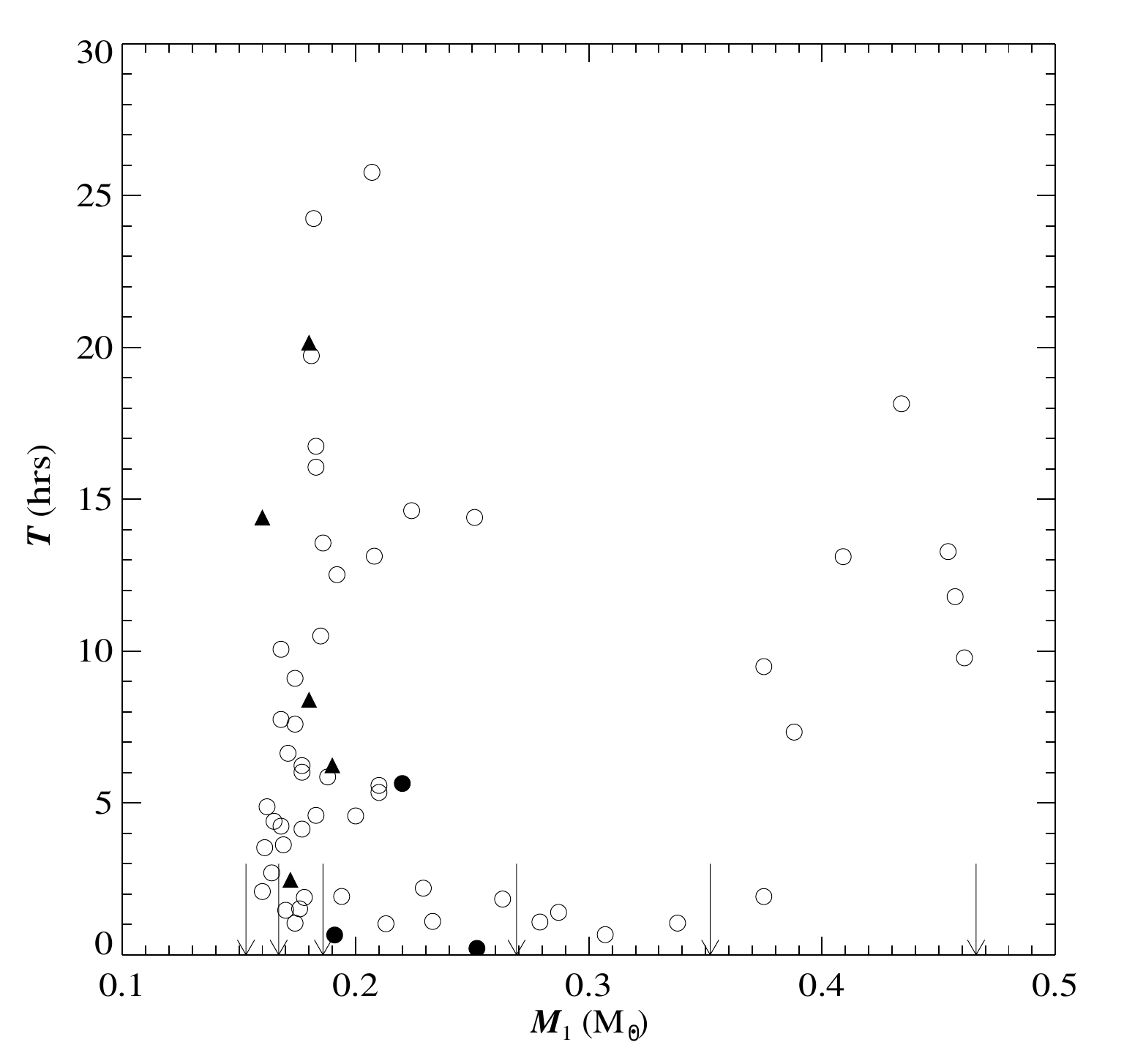}
\caption{The $M_1$ - $\period$ distribution of the ELM sample (circles) and the known WD-NS binaries (triangles). The three eclipsing systems in the ELM sample with known $M_2$ are shown as filled circles, and the masses of the ELM WDs without detected RV variations are shown by the arrows. From $M_1$ and $\period$ alone, the two populations are indistinguishable.}
\label{fig:Porb-M1}
\end{center}
\end{figure}

For each LMWD in the ELM sample, spectroscopy provides $\period$, the primary WD mass $M_1$, and the projected orbital velocity $K=v \sin i$. Assuming circular orbits, we can write:
\begin{equation}
	\frac{(M_2 \sin i)^3}{\left(M_1+M_2\right)^2} = \frac{\period}{2\pi G} K^3, \label{eq:massfunc}
\end{equation}
where the right side is the mass function $\mf$. The companion mass, $M_2$, is minimized for an edge-on orbit ($i = 90\degree$). Because of this dependence on $i$, the nature of the companion cannot usually be determined based on $\mf$ alone. Figure~\ref{fig:Porb-M1} shows that the population of LMWDs with pulsar companions occupies the same region in $M_1 - \period$ space as those with WD companions. Therefore, barring rare circumstances such as eclipsing systems, individual LMWDs with NS companions cannot be identified from optical observations alone.

The ELM sample is now large enough that the $M_2$ distribution and NS companion fraction can be constrained statistically. 
We have developed a probabilistic model to infer parameters of an assumed form for the $M_2$ distribution. Our method is similar to that employed by \citet{ozel12} and \citet{kiziltan13} to describe the mass distribution of NSs in binaries using post-Keplerian parameters. We focus on the following questions: Can the companion population be modeled using a simple description of $M_2$? How does the $M_2$ distribution compare to predictions from population synthesis simulations?  What is the rate of LMWD-NS binaries implied by our model? What are the resulting distributions of NS probabilities for individual systems in the ELM sample? 

To answer these questions, we build the mathematical framework (Section 2), then test our resulting model (Section 3). We apply our model to the ELM sample (Section 4) before concluding (Section 5).

\section{Building our model}
We construct a statistical model to constrain a parametric model for the distribution of LMWD companion masses, $p(M_2 \given \bs{\theta})$.\footnote{We represent vectors or sets of parameters or quantities by bold symbols.} For each system, we assume we have $K$, $T$, and $M_1$, and therefore know $\mf$. We wish to derive posterior constraints on the model parameters, $\bs{\theta}$, which describe the distribution of companion masses, $p(M_2\given \bs{\theta})$, given the set of observed mass functions, $\bs{m_f}$, by deconvolving the $\mf$ distribution from the unobserved inclinations. Using Bayes' rule,
\begin{equation}
    p(\bs{\theta} \given \bs{\mf}) = \frac{1}{\mathcal{Z}}~p(\bs{\mf} \given \bs{\theta})~p(\bs{\theta}),
\end{equation}
where $p(\bs{\mf} \given \bs{\theta})$ is the likelihood, $p(\bs{\theta})$ is the prior on parameters $\bs{\theta}$, and the evidence integral, $\mathcal{Z}$, is a constant that depends only on the data. The likelihood, $p(\bs{\mf} \given \bs{\theta})$, can be split into a product over the likelihoods of individual systems:
\begin{equation}
p(\bs{\mf} \given \bs{\theta}) = \prod_j p(\mf \given \bs{\theta}),
\end{equation}
where the product is over each of the $j$ systems.  This marginal likelihood involves integrals over the unobserved quantities $i$ and $M_2$,
\begin{align}
    p(\mf \given \bs{\theta}) &= \int_0^\infty dM_2 \int_0^{\pi/2} di  \nonumber \\
      & \qquad {} \times p(\mf \given M_1, M_2, i)~p(M_2 \given \bs{\theta})~p(i).
\end{align}
We neglect observational uncertainties in $\mf$ and $M_1$,\footnote{The fractional uncertainties in these quantities are small, $\sigma_x / x \sim 0.05-0.1$ \citep{gianninas14}.} and assume the inclination angles are isotropically distributed:
\begin{equation}
	p(\mf \given M_1, M_2, i) = \delta \left[\mf - f(M_1, M_2, i) \right],
\end{equation}
where
\begin{equation}
	f(M_1, M_2, i) = \frac{(M_2 \sin i)^3}{(M_1 + M_2)^2}
\end{equation}
and
\begin{equation}
p(i) = \sin i.
\end{equation}
For now, we do not specify a parametric form for the companion mass distribution, $p(M_2 \given \bs{\theta})$. With the above assumptions, the marginal likelihood integral is:
\begin{align}
    p(\mf \given \bs{\theta}) &= \int_{0}^\infty dM_2 ~p(M_2 \given \bs{\theta})  \nonumber \\
    & \qquad {} \times \int_0^{\pi/2} di ~\sin i ~ \delta \left[g(M_1,M_2,i) \right]\label{eq:delta},
\end{align}
where
\begin{equation}
	g(M_1,M_2,i) = \mf - \frac{M_2^3}{(M_1+M_2)^2}\sin^3 i.
\end{equation}
The inner integral (over $i$) has the form:
\begin{equation}
    \int dx~F(x)~\delta \left[ G(x) \right] = \sum_j \frac{F(x^*_j)}{|G'(x^*_j)|},
\end{equation}
where the sum is over the roots, $x^*_j$, of the function $G(x)$. The root, $i^*$, and derivative of the argument of the delta function in \eqn\ref{eq:delta} are: 
\begin{align}
	\sin i^* &= \frac{ \left[\mf(M_1+M_2)^2 \right]^{1/3}}{M_2}, \\
	\frac{\partial g}{\partial i}\bigg\rvert_{i^*} &= \frac{3M_2^3}{(M_1+M_2)^2} \sin^2 i^* \sqrt{1 - \sin^2 i^*}.
\end{align}
We may rewrite the marginal likelihood as:
\begin{align}
	p(\mf \given \bs{\theta}) &= \int_{0}^\infty dM_2~p(M_2 \given \bs{\theta})~\sin i^* \left(\frac{\partial g}{\partial i}\bigg\rvert_{i^*}\right)^{-1}\\
	&= \int_{M_{2,{\rm min}}}^\infty dM_2~p(M_2 \given \bs{\theta})~h(M_2, \mf, M_1). \label{eq:fullm2}
\end{align}
The bottom bound in the integral in \eqn\ref{eq:fullm2} is set by the minimum companion mass for which the integrand is real, $M_{2,{\rm min}}$, determined by setting $i=90\degree$ in \eqn\ref{eq:massfunc} and solving for $M_2$, and

\begin{equation}
h(M_2, \mf, M_1) = \frac{(M_1+M_2)^{4/3}}{3\ \mf^{1/3}M_2\sqrt{M_2^2 - \left[ \mf(M_1+M_2)^2 \right]^{2/3}}}.
\end{equation}

\subsection{Our Model} \label{sec:experiments}
We must now choose a functional form for the companion mass distribution, $p(M_2\given \bs{\theta})$. We use a two-component Gaussian mixture model. We truncate the distributions using physically motivated bounds: the WD component is restricted to $M_2\in [0.2,\wdupper]~\Msun$ and the NS component is restricted to $M_2\in [1.3,2.0]~\Msun$. We then have:
\begin{align}
	p(M_2 \given \bs{\theta}) &= \left[ (1-f_{\rm NS})~p_{\rm WD} + f_{\rm NS}~p_{\rm NS} \right], 
\end{align}
where $f_{\rm NS}$ is the NS fraction and
\begin{align}
	p_{\rm WD} &= \mathcal{N}(M_2 \given \mu_{\rm WD}, \sigma^2_{\rm WD}); ~0.2 < \frac{M_2}{\Msun} < \wdupper, \\
	p_{\rm NS} &= \mathcal{N}(M_2 \given \mu_{\rm NS}, \sigma^2_{\rm NS}); ~1.3 < \frac{M_2}{\Msun} < 2.
\end{align}
$\mathcal{N}$ is the (truncated, but properly normalized) normal distribution with mean $\mu$ and variance $\sigma^2$; the distributions are limited to the ranges specified. To reduce the number of parameters in our model we fix $\mu_{\rm NS}$ and $\sigma_{\rm NS}$
to:
\begin{align}
	\mu_{\rm NS} &= 1.4~\Msun, \\
	\sigma_{\rm NS} &= 0.05~\Msun,
\end{align}
as some NSs in binaries may be somewhat more massive than the canonical NS mass of 1.35 \Msun~\citep{kiziltan13,smedley14}.

The probability of any particular WD having a NS companion, $P_{\rm NS}$, can be computed for a given set of parameters for the $M_2$ distribution:
\begin{equation}
P_{\rm NS} = \frac{\int_{M_{2,{\rm min}}}^{\infty} dM_2~ f_{\rm NS}~ p_{\rm NS}~ h(M_2, \mf, M_1)}{p(\mf \given \bs{\theta})}. \label{eq:P_NS}
\end{equation}

Our companion mass model parameters are then $\bs{\theta} = (\mu_{\rm WD}, \sigma_{\rm WD}, f_{\rm NS})$. For $\mu_{\rm WD}$, we use a uniform prior from $0.2-1.0~\Msun$; for $\sigma_{\rm WD}$, we use a logarithmic (scale-invariant) prior over the range $0.02-2.0~\Msun$. Finally, we use a uniform prior over the dimensionless $f_{\rm NS}$ from $0-1$. The model parameters are summarized in Table~\ref{tbl:parameters}.

\renewcommand{\arraystretch}{1.405}
\begin{deluxetable}{ccccc}
	\tablecaption{Model Results \label{tbl:parameters}}

	\tablehead{
		\multicolumn{2}{c}{} &
		\colhead{$\mu_{\rm WD}$} & 
		\colhead{$\sigma_{\rm WD}$} &
		\colhead{$f_{\rm NS}$} \\
		\colhead{} &
		\colhead{} &
		\colhead{[\Msun]} &
		\colhead{[\Msun]} &
		\colhead{}
	}

	\startdata
		\multicolumn{2}{c}{\multirow{2}{*}{Priors}} & $\mathcal{U}(0.2, 1)$ & $\propto \sigma^{-1}$  & $\mathcal{U}(0, 1)$ \\
		\multicolumn{2}{c}{} & & $(0.02 < \sigma/\Msun < 2.0)$ & \\
		\cutinhead{Test Cases}
		\multirow{2}{*}{Test 1} & True & 0.7 & 0.2 & 0 \\
		 & MAP & 0.72 & 0.20 & 0.0 \\
		 \hline
		\multirow{2}{*}{Test 2} & True & 0.7 & 0.2 & 0.10 \\
		 & MAP & 0.74 & 0.19 & 0.11 \\
		 \hline
		\multirow{2}{*}{Test 3} & True & \nodata & \nodata & 0.10 \\
		 & MAP & 0.63 & 0.52 & 0.14 \\
		 \hline
		\multirow{2}{*}{PCEB} & True & \nodata & \nodata & 0 \\
		 & MAP & 0.58 & 0.16 & 0.0 \\
		 \cutinhead{ELM Sample}
		 & MAP & 0.74 & 0.24 & 0.0
	\enddata

	\tablecomments{Parameter information for the form of the $M_2$ distribution used in the tests described in Section~\ref{sec:tests}. $\mathcal{U}$ is the uniform distribution. We additionally fix the NS mass distribution: $\mu_{\rm NS} = 1.4~\Msun$ and $\sigma_{\rm NS} = 0.05~\Msun.$}\

\end{deluxetable}

We use a Markov Chain Monte Carlo algorithm \citep{goodman10} to draw samples from the posterior distribution, $p(\mu_{\rm WD}, \sigma_{\rm WD}, f_{\rm NS} \given \bs{m}_f, \bs{M}_1)$.\footnote{Our model uses {\tt emcee}, implemented in \texttt{Python} \citep{foremanmackey13}.} The algorithm uses an ensemble of individual ``walkers'' to naturally adapt to the geometry of the parameter-space being explored. We run the walkers for a burn-in period of 500 steps starting from randomly drawn initial conditions (sampled from the priors in Table~\ref{tbl:parameters}). We then re-initialize the walkers from their positions at the end of this run and run again for 1000 steps. We remove the burn-in samples to eliminate any effects due to our choice of initial conditions. 

\section{Testing Our Model} \label{sec:tests}
We test the performance of this Gaussian mixture model on four separate data sets: three mock data sets and a sample of SDSS post-common-envelope binaries \citep[PCEBs;][]{nebot11}. 
Each of the 100 systems in our three mock data sets is generated by computing a $\mf$ from a random $M_1$ (drawn from a uniform distribution, $\mathcal{U}(0.2,0.4)~\Msun$), $M_2$ (from the distributions described below), and $i$ (from an isotropic distribution). 
We apply the same Gaussian mixture model to all four tests to infer the parameters of the WD mixture component and $f_{\rm NS}$.

\begin{figure*}[h!]
\begin{center}
\includegraphics[width=0.95\textwidth]{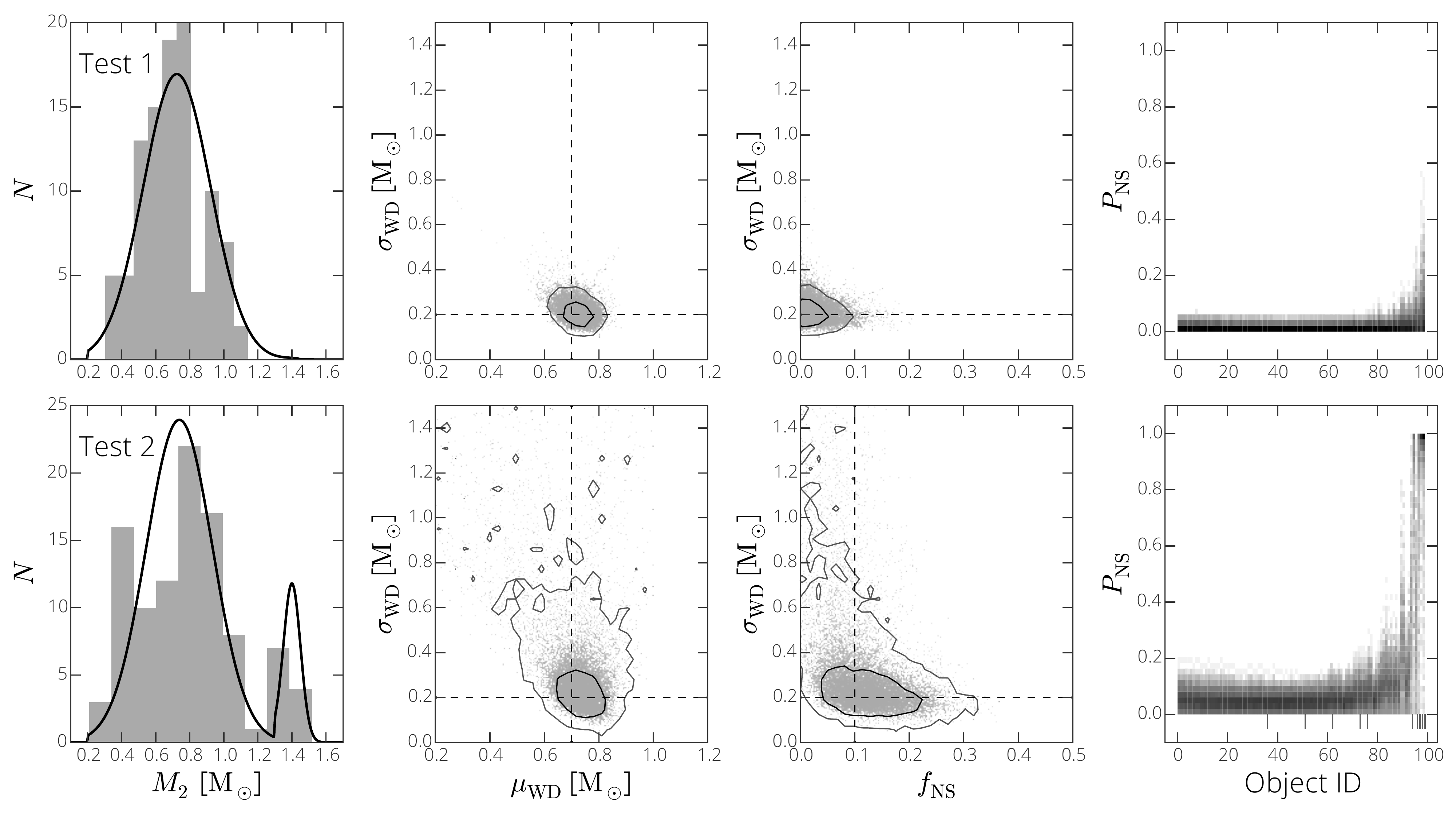}
\caption{Results from testing the first two mock data sets described in Section~\ref{sec:tests}. The left-most panels show the companion masses (gray histogram) randomly drawn from each of our test distributions and our MAP models (black line). Panels in the second and third columns show samples from the posterior distributions of $\mu_{\rm WD}$ and $\sigma_{\rm WD}$ and $f_{\rm NS}$. Contours designate the 68\% and 95\% confidence levels. Dashed lines in these panels show the true values from which the sample systems were drawn. The fourth panel shows individual mock LMWD systems (ordered by increasing $\mf$) and their corresponding $P_{\rm NS}$ distribution. Tick marks along the bottom indicate inputed LMWD-NS systems.}
\label{fig:tests_1_2}
\end{center}
\end{figure*}

\subsection{Test 1: Single Gaussian (WD)} \label{sec:exp1}
We first generate companion masses by drawing from a single, truncated Gaussian with the parameters given in Table~\ref{tbl:parameters}. This mock sample contains no NSs. In the top row of Figure~\ref{fig:tests_1_2}, the left-most panel shows that our model finds a maximum a posteriori (MAP) $M_2$ distribution (black line) that qualitatively matches the input distribution (gray histogram). The second and third panels show samples from the posterior distributions and contours containing 68\% and 95\% of the samples for our three model parameters. The input values (dashed lines) lie cleanly within the inner contour in both panels, although $f_{\rm NS}$ has a tail up to $\approx$10\%.

Equation~\ref{eq:P_NS} gives the probability of an individual system hosting a NS. Using posterior samples, we can determine the distribution of $P_{\rm NS}$ for each system. The right-most panel in Figure~\ref{fig:tests_1_2} includes all the individual systems, ordered by $\mf$, and shows the distributions of $P_{\rm NS}$ for each. For most systems, there is negligible probability above $P_{\rm NS}\sim 5\%$.

\subsection{Test 2: Two Gaussians (WD + NS)} \label{sec:exp2}
We use the same Gaussian distribution to generate companion masses for the WDs but add a NS component with $f_{\rm NS} = 10\%$. The bottom row of Figure~\ref{fig:tests_1_2} shows that our model again recovers the input values for $\mu_{\rm WD}$ and $\sigma_{\rm WD}$. 
Importantly, the third panel shows that our model also recovers $f_{\rm NS}$, although the posterior shows a substantial tail toward higher $f_{\rm NS}$. Tick marks in the right-most panel of Figure~\ref{fig:tests_1_2} indicate ``true" NSs in our mock data. Our model correctly assigns high $P_{\rm NS}$ to roughly half of these. However, many systems with NS companions have inclinations too low to be statistically differentiated from those with WD companions.

\subsection{Test 3: Uniform (WD) + Gaussian (NS)} \label{sec:exp3}
We generate companion masses for the WDs by sampling from a uniform distribution over $[0.2,1.2]~\Msun$, again with $f_{\rm NS}=$10\%. The top row of Figure~\ref{fig:tests_3_4} shows the results. The posterior distribution in the second panel indicates that $\mu_{\rm WD}$ and $\sigma_{\rm WD}$ are not well constrained. The preference for larger $\sigma_{\rm WD}$ is expected, as the model flattens the Gaussian model distribution to match it with the input uniform distribution. Interestingly, the third panel shows that despite having a non-Gaussian input distribution for $M_2$, and a poorly constrained $\sigma_{\rm WD}$, our model still recovers $f_{\rm NS}$ approximately as accurately as in Test 2. Furthermore, the fourth panel of Figure~\ref{fig:tests_3_4} demonstrates that our model effectively identifies which LMWDs host NS companions.

\subsection{Test 4: PCEBs} \label{sec:PCEB}
\begin{figure*}[h!]
\begin{center}
\includegraphics[width=0.95\textwidth]{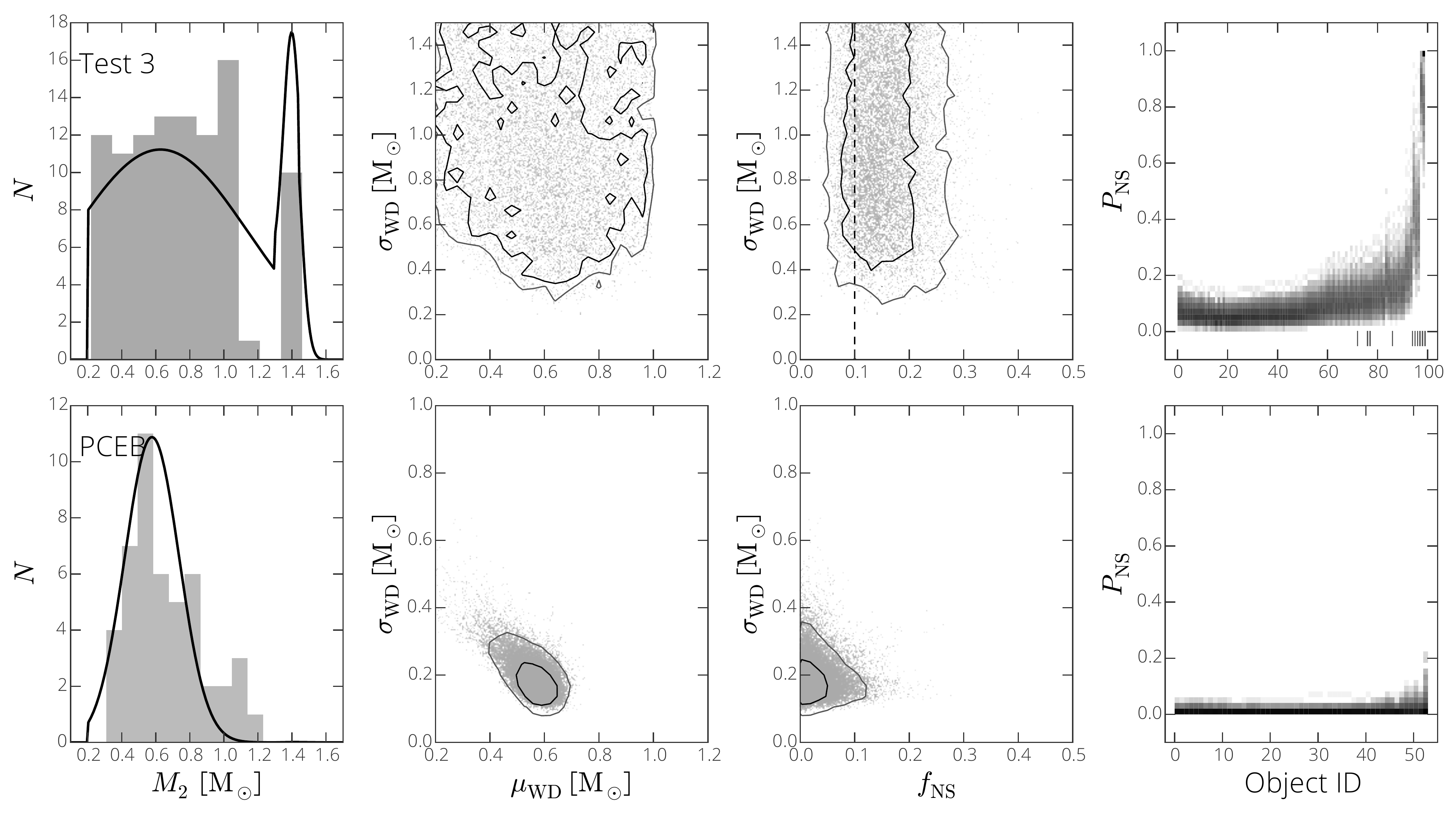}
\caption{ The results of our model when applied to our third mock data set and the SDSS PCEB sample. The panels are same as those in Figure~\ref{fig:tests_1_2}.}
\label{fig:tests_3_4}
\end{center}
\end{figure*}

PCEBs are composed of WDs in close orbits with main-sequence companions. The \citet{nebot11} sample of 54 SDSS PCEBs, which have precisely determined $K$, $\period$, and masses for the main-sequence companions, are an ideal test sample for our model. Our model uses these parameters to try and recover the PCEBs WD mass distribution, which we can then compare to the spectroscopically determined WD masses.
Our MAP distribution (black line) is shown in the left-most panel in the bottom row of Figure~\ref{fig:tests_3_4}. Our model qualitatively recovers the true $M_{\rm WD}$ distribution (gray histogram). The third panel shows that the posterior $f_{\rm NS}$ distribution is very low, as expected since there are no NS companions in the PCEB sample. This is further illustrated in the right-most panel, where every PCEB in the sample has low $P_{\rm NS}$ values.

\section{Applying our model}
\subsection{The ELM Sample}
The ELM WD Survey is based on the Hypervelocity Star Survey \citep{brown06}, and includes previously identified SDSS LMWDs \citep{eisenstein06,liebert04}. Objects are chosen for spectroscopic follow-up based on their $ugr$ colors, and this choice is independent of the mass and nature of any putative companions. Therefore, at least with regard to $i$ and $M_2$, the population is unbiased.

The ELM WD sample includes 55 systems with RV variations fit to orbital solutions, which provide precise measurements of $\period$ and $K$. WD masses in these systems are derived from fits to spectroscopic templates, which are generally precise to $\approx$10\% \citep{gianninas14}. The masses of cool LMWDs may suffer somewhat from inaccuracies in the one-dimensional WD atmospheric models \citep{tremblay13}. However, since this should only affect the coolest WDs in the ELM sample, we expect any impact on our results to be minor.

Three systems are eclipsing binaries, with known companion masses: NLTT 11748 \citep[$M_2=0.72~\Msun$;][]{kaplan14}, SDSS J065133.3$+$284423.3 \citep[$M_2=0.50~\Msun$;][]{brown11b}, and SDSS J075141.2$-$014120.9 \citep[$M_2=0.97~\Msun$;][]{kilic14}. For these systems, the likelihood reduces to:
\begin{equation}
p(\mf \given \theta) = (1-f_{\rm NS}) \mathcal{N}(M_2^* \given \mu_{\rm WD}, \sigma^2_{\rm WD}),
\end{equation}
where $M_2^*$ is the mass of the WD companion. 

The other six ELM systems show no evidence of orbital motion, with RV upper limits of $\approx$20-50 km s$^{-1}$. Some of these systems may be in low $i$ binaries with RVs below the detection limit, or may have $\period\approx24$ hr, which is difficult to measure \citep{ELMV}. These LMWDs could also have companions at systematically longer $\period$, resulting in orbital velocities below the detection limit. We do not include these systems in our analysis.

\begin{figure*}[h!]
\begin{center}
\includegraphics[width=0.95\textwidth]{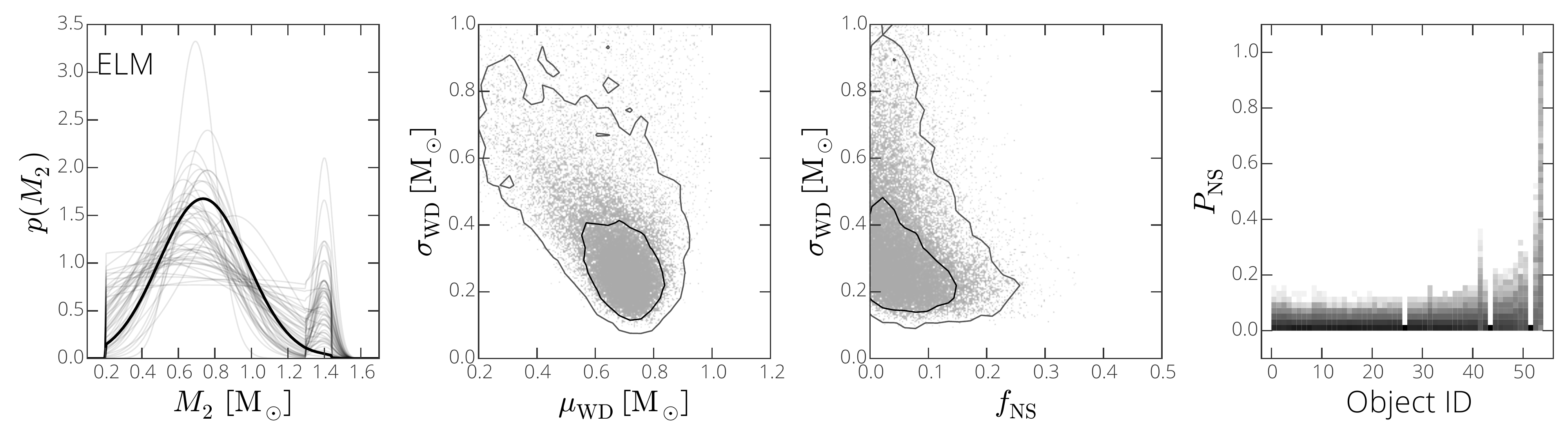}
\caption{Results from applying our model to the ELM WDs. 
The panels are the same as in Figures~\ref{fig:tests_1_2} and \ref{fig:tests_3_4}. 
The left-most panel shows both the MAP $M_2$ distribution (solid black) and random samples from the posterior (gray lines). 
The three systems in the right-most panel with all $P_{\rm NS} = 0\%$ are the eclipsing systems with measured $M_2$.
}
\label{fig:ELM_post}
\end{center}
\end{figure*}

\subsection{Results and Discussion}
The results from applying our model to the ELM sample are shown in Figure~\ref{fig:ELM_post}. The MAP model gives $\mu_{\rm WD} = 0.74~\Msun$, $\sigma_{\rm WD} = 0.24~\Msun$, and $f_{\rm NS} = 0\%$. The marginal posterior over $\mu_{\rm WD}$ and $\sigma_{\rm WD}$ has a tail toward larger $\sigma_{\rm WD}$, which could indicate that the true WD distribution may not be exactly Gaussian.

It is interesting that the best-fit Gaussian for the companions to the ELM WDs is similar to that of the population of single hydrogen-atmosphere WDs in SDSS, with a mean of 0.6 $\Msun$ \citep{kleinman13}. Our MAP variance is significantly larger: $\sigma \approx 0.26~\Msun$, compared to $\sigma \approx 0.1~\Msun$, possibly due to past mass transfer phases increasing the masses of the unseen primary WDs.

The low combined mass in these systems indicates that, although several of them will merge within a Hubble time \citep{ELMV}, the majority of the ELM systems are unlikely to be type Ia SN progenitors. However, we cannot rule out the possibility that some individual LMWD binaries may be massive enough to produce type Ia SNe \citep{justham09}.

Our posterior distributions further suggest that the companions to LMWDs have predominantly CO cores. This is in contrast to population synthesis models, which suggest that LMWDs should predominantly have He-core WD companions \citep{toonen12}. With a larger sample, a more sophisticated LMWD companion model could place quantitative constraints on population synthesis predictions.

The third panel in Figure~\ref{fig:ELM_post} shows a $f_{\rm NS}$ strongly peaked toward 0\%. However, there is a significant tail toward higher NS probabilities. Our model indicates $f_{\rm NS} <16\%$ at the 68\% confidence level, in agreement with independent constraints from \citet[][$f_{\rm NS}<18\pm5$\%]{vLeeuwen07} and \citet[][$f_{\rm NS}<10\substack{+4 \\ -2}~\%$]{agueros09b}, both based on radio non-detections of LMWD companions.

The right-most panel in Figure~\ref{fig:ELM_post} indicates there are two LMWDs with substantial $P_{\rm NS}$: SDSS J081133.6$+$022556.8 and J174140.5$+$652638.7. However, the X-ray non-detection of SDSS J174140.5$+$652638.7 suggests its companion is unlikely to be a NS \citep{kilic14}. Searches for radio and X-ray emission from SDSS J081133.6$+$022556.8 are on-going. We note that the $P_{\rm NS}$ distributions in each of our samples show a trend such that systems with higher $\mf$ have higher $P_{\rm NS}$ values. These high $\mf$ systems are therefore ideal targets to search for NS companions to LMWDs.

\section{Conclusions}
We have developed a statistical model to infer the companion mass distribution for a sample of single-line, spectroscopic binaries. This model can be applied to any such sample with measured $M_1$ and $\mf$. When tested on three separate mock data sets with unseen WD and NS companions to LMWDs, our model recovers the input parameters. Even when the companion mass distribution is not drawn from a Gaussian distribution, our model still infers the input NS fraction to within a few percent. We further apply our model to the SDSS PCEBs \citep{nebot11}, and our model qualitatively recovers the independent, spectroscopically measured $M_{\rm WD}$ distribution. 

We applied our model to the set of LMWDs from the ELM WD survey. The resulting posterior distribution is qualitatively similar to our two-component Gaussian test case, suggesting that the companion mass distribution to the LMWDs in the ELM sample is well-described by our model. Our model returns a MAP $\mu_{\rm WD} = 0.74\pm0.24\ \Msun$, suggesting that a majority of ELM WDs have CO-core WD companions. This is in contrast to predictions from population synthesis models, which find that the dominant companion population should be He-core WDs \citep[e.g.,][]{toonen12}. Our model further indicates that the fraction of ELM WDs with NS companions is consistent with 0\%, but could be as high as $\approx$16\% (within 1-$\sigma$). Finally, our model identifies the LMWD SDSS J081133.6$+$022556.8 as having the highest median probability of hosting a NS companion.

To determine the probability of any particular LMWD hosting a NS, we make our model posteriors publicly available on fig{\bf share}.\footnote{\url{http://dx.doi.org/10.6084/m9.figshare.1206621}} We further provide a {\tt Python} script that calculates $P_{\rm NS}$ and the mass distribution for a WD companion for any LMWD with a measured $M_1$ and $\mf$. This script can be applied to newly discovered LMWDs as well as those already in the ELM sample.

There are several ways in which our model can be expanded. By modeling photometric variability, \citet{hermes14} recently constrained the inclination of 20 LMWDs in the ELM sample; we could include these constraints. Furthermore, our model can place tighter constraints on $f_{\rm NS}$ by factoring in radio and X-ray non-detections. We plan to develop our method to quantitatively compare our model to the results of population synthesis codes, potentially constraining the formation of LMWDs.

\acknowledgements
The authors thank David Hogg, DJ D'Orazio, and Josh Peek for useful discussions, and the organizers of the \emph{AstroData Hack Week} (2014). We are grateful to the anonymous referees for comments that helped improve this paper. MAA acknowledges support provided by the NSF through grant AST-1255419. APW is supported by a NSF Graduate Research Fellowship under grant No.\ 11-44155. This research made use of Astropy, a community-developed core \texttt{Python} package for Astronomy \citep{astropy13}. \\

\bibliographystyle{apj}

%\bibliography{refs}
  
%\bibitem[Goodman~\&\ Weare(2010)]{goodman10}
%Goodman,~J. \& Weare,\ J.
%2010, Comm.\ App.\ Math.\ Comp.\ Sci., 5, 65

\end{document}